\documentstyle[eqsecnum,aps,epsf]{revtex}
\begin{document}
\draft

\newcommand {\beq}{\begin{eqnarray}}
\newcommand {\eeq}{\end{eqnarray}}
\newcommand {\be}{\begin{equation}}
\newcommand {\ee}{\end{equation}}
\newcommand{\Gmu}{\gamma^{\mu}}
\newcommand{\Gnu}{\gamma^{\nu}}
\newcommand{\gmu}{\gamma_{\mu}}
\newcommand{\gnu}{\gamma_{\nu}}
\newcommand{\bg}{\mbox{\boldmath $\gamma$}}
\newcommand{\gfour}{\gamma_4}
\newcommand{\del}{\partial}
\newcommand{\k}{\mbox{\boldmath $k$}}
\newcommand{\q}{\mbox{\boldmath $q$}}
\newcommand{\p}{\mbox{\boldmath $p$}}
\newcommand{\wn}{\omega_n}
\newcommand{\wm}{\omega_m}
\newcommand{\La}{{\Lambda_{\scriptsize{\mbox{QCD}}}}}
\newcommand{\CC}{\langle \bar{q} q \rangle}

\topmargin=0cm

\title{Dressed Quark Propagator at Finite Temperature 
       in the Schwinger-Dyson approach with the Rainbow Approximation\\
       - exact numerical solutions and their physical implication -}

\author{Takashi Ikeda\footnote{\tt ikeda@nt.phys.s.u-tokyo.ac.jp}}

\address{Department of Physics, University of Tokyo, Tokyo 113-0033, Japan}

\date{\today}

\maketitle

\begin{abstract}

 The Schwinger-Dyson equation for the quark in the rainbow 
approximation at finite temperature ($T$) is solved numerically without 
introducing any ansatz for the dressed quark propagator. 
The dymanical quark mass-function and the wave-function 
renormalization are found to have non-trivial dependence on 
three-momentum, Matsubara-frequency and temperature.
The critical temperature of the chiral phase transition ($T_c$) 
and the $T$-dependence of the quark condensate are highly affected 
by the wave-function renormalization. We found that 
$T_c \simeq 155$ MeV which is consistent with the result of the
finite temperature lattice QCD simulation.
It is also found that the system is not a gas of 
free quarks but a highly interacting system
of quarks and gluons even in the chirally symmetric phase.

\end{abstract}

%%%%%%%%%%%%%%%%%%%%%%%%%%%%%%%%%%%%%%%%%%%%%%%%%%%%%%%%%%%%%%%%%%%%%%%%%%
%                         SECTION.1
%%%%%%%%%%%%%%%%%%%%%%%%%%%%%%%%%%%%%%%%%%%%%%%%%%%%%%%%%%%%%%%%%%%%%%%%%%

\section{INTRODUCTION}

 The vacuum of the quantum chromodynamics (QCD) is believed to
undergo a phase transition to the chirally symmetric and deconfinement
phase at high temperature and/or high baryon-density due to the 
asymptotic freedom and the plasma screening. Such a new state of 
matter is called the quark-gluon plasma (QGP) and is expected to 
be produced  in the on-going Relativistic Heavy-Ion Collider (RHIC) 
at BNL and in the future Large Hadron Collider (LHC) at CERN
\cite{QMproceedings}.
Furthermore, in the core of the neutron stars,
cold but high-density quark matter with a color
superconductivity may be realized as has been extensively
discussed in recent years
\cite{Rajagopal:2000wf}.

In relation to QGP, it is of great importance to theoretically 
understand the phase structure of QCD at finite temperature and density
\cite{phase}.
The lattice QCD simulation is one of the  most powerful methods
for this purpose since it is based on the first principle QCD.
Although the finite density simulation is still not available, the
phase transition at finite $T$ in full QCD is extensively studied on 
the lattice
\cite{lattice}.
Another method, which could be potentially very useful to study the QCD 
phase structure, is the Schwinger-Dyson (SD) approach
\cite{SDreview,quarkSD1,quarkSD2}. 
This method is based on the continuum field theory and has been applied 
for analysing the spontaneous chiral symmetry breaking and confinement
in the vacuum. Exact SD equations are infinitely coupled integral 
equations and their solutions yield the n-point Euclidean Green's 
functions. In the practical applications, one needs to truncate the 
coupled SD equations by introducing some ans\"{a}tze. 

 When one studies the spontaneous chiral symmetry breaking at zero temperature 
and density in the SD approach, one starts with the dressed quark 
propagator in the Euclidean space
\cite{SDreview}, 
\beq 
S(p) &=& \frac{1}{i \slash \hspace{-5.8pt} p A(p^2) + B(p^2)} \nonumber \\
     &\equiv& -i\slash \hspace{-5.8pt} p \sigma_A(p^2) + \sigma_B(p^2) .
\label{eq:T0propa}
\eeq
Here, $A(p^2)$ is the wave-function renormalization and $B(p^2)$ is 
the quark mass-function. If one finds a solution $B(p^2) \neq 0$ 
in the chiral limit, it implies the spontaneous 
breaking of the chiral symmetry
\cite{QWFR}.
If the free gluon propagator in the Landau-gauge and the bare 
quark-gluon vertex are used in the 1-loop SD equation for the quark, 
$A(p^2) \equiv 1$ holds exactly
\cite{JM_91,PS}. 
However, this is not the case for other models of the gluon propagator 
and/or the quark-gluon vertex.
The SD equation for the quark is obtained by the extremum condition of 
the effective potential under the variation of the dressed quark 
propagator $S$
\cite{CJT}.
There are two distinct solutions of $S$ in the chiral limit
corresponding to the Wigner phase and the Nambu-Goldstone phase.
The true ground state should be determined by calculating the effective 
potential in those two phases.

 The introduction of temperature or baryon-density to Euclidean QCD 
reduces the $O(4)$ symmetry to $O(3)$. As a consequence, the dressed 
quark propagator in the imaginary-time formalism has a general form 
\cite{SDreview},
\beq 
S(\p,\wn) &=& \frac{1}{i \bg \cdot \p A(\p^2,\wn^2) 
          + i \gfour \wn C(\p^2,\wn^2) + B(\p^2,\wn^2)} \nonumber \\
          &\equiv& -i\bg \cdot \p \sigma_A(\p^2,\wn^2) - i \gfour \wn 
          \sigma_C(\p^2,\wn^2) + \sigma_B(\p^2,\wn^2).
\label{eq:propa}
\eeq
Here $\wn = (2n + 1)\pi T + i \mu \, (n \in Z)$
is the Matsubara-frequency for the quark.
$T$ and $\mu$ are the temperature and the quark chemical 
potential respectively. $A(\p^2,\wn^2)$ and $C(\p^2,\wn^2)$ represent
the quark wave-function renormalization, and $B(\p^2,\wn^2)$ is the quark
mass-function. For $T \neq 0$ and/or $\mu \neq 0$,
$A(\p^2,\wn^2)=C(\p^2,\wn^2)=1$ does NOT hold even if the free gluon 
propagator in the Landau-gauge and the bare quark-gluon vertex are used in 
the 1-loop SD equation.

 The purpose of this paper is to obtain a dressed quark propagator 
at $T \neq 0$ by solving the 1-loop 
SD equation for the quark without 
any ansatz for the functions $A,C$ and $B$, and to investigate their 
physical implications. The exact numerical solutions of $A,C$ and $B$ have 
explicit ($\p^2,\wn^2$) dependences. Although such functions have been 
obtained in a simple model where the SD equation can be solved algebraically
\cite{SDreview},
the ($\p^2,\wn^2$) dependence of $A$ and $C$ for realistic models
such as QCD-like theory
\cite{KMT}
has not been known. Non-trivial ($\p^2,\wn^2$) dependence of the 
functions $A$ and $C$ to the nonperturbative phenomena of QCD in 
the Wigner phase will be also discussed.

 The paper is organized as follows. In Sec.~II.~A, the SD equation
and the CJT effective potential for the quark are summarized.
Then a model for the gluon propagator, the quark-gluon vertex 
and the QCD coupling constant is introduced at $T = 0$.
In Sec.~II.~B, they are extended to $T \neq 0$. In Sec.~III, 
basic parameters of the model introduced in Sec.~II are determined by 
the condition $f_\pi$ = 93 MeV at $T=0$. Then the quark wave-function 
renormalization $A(\p^2,\wn^2)$ and $C(\p^2,\wn^2)$ and the quark 
mass-function $B(\p^2,\wn^2)$ are obtained self-consistently at $T \neq 0$
by solving the SD equation with simultaneous iteration. By calculating 
the effective potential corresponding to those solutions, the true
ground state is determined at given $T$. The critical 
temperature $T_c$ at which the chiral symmetry is restored is also obtained. 
The effect of the ($\p^2,\wn^2$) dependence of $A$ and $C$ on the 
chiral phase transition is
examined. In Sec.~IV conclusions and discussions are given. In Appendix, 
the explicit forms of the integral kernels which appear in 
the SD equation for the quark are given. In this paper, 
massless (u,d)-quarks are considered, and 
all calculations are done in the imaginary-time formalism
\cite{ITF}.

%%%%%%%%%%%%%%%%%%%%%%%%%%%%%%%%%%%%%%%%%%%%%%%%%%%%%%%%%%%%%%%%%%%%%%%%%%
%                         SECTION.2
%%%%%%%%%%%%%%%%%%%%%%%%%%%%%%%%%%%%%%%%%%%%%%%%%%%%%%%%%%%%%%%%%%%%%%%%%%

\section{SD EQUATION AND EFFECTIVE POTENTIAL}

\subsection{SD eq. for the quark at $T = 0$ in the rainbow
            approximation and the CJT effective potential}

 At $T = 0$, the SD equation for the quark in the chiral 
limit is expressed as,
\be 
S^{-1}(p) =  S^{-1}_{(0)}(p) + \Sigma(p) \, . 
\label{eq:T0sdeq}
\ee
Here, $S(p)$ and $S_{(0)}(p)$ are the dressed quark propagator in 
eq. (\ref{eq:T0propa}) and the free quark propagator
$1/i\slash \hspace{-5.8pt} p$ in the chiral limit, respectively.
$\Sigma(p_{\mu})$ represents the self-energy of the quark.
Eq. (\ref{eq:T0sdeq}) is derived by the extremum condition for the CJT 
effective potential $V[S]$ under the variation of $S(p)$
\cite{CJT};
\beq
V[S] &=& V_1[S] + V_2[S] ,\\
V_1[S] &=& \int \frac{d^4p}{(2\pi)^4} \mbox{tr} \left\{ \ln \left[
       S_{(0)}^{-1}(p)S(p) \right] - S_{(0)}^{-1}(p)S(p) + 1 \right\} ,\\
V_2[S] &=& - \frac{1}{2} \int \frac{d^4p}{(2\pi)^4} \left\{ \mbox{tr}
       \left[ \Sigma(p) S(p) \right] \right\} .
\eeq
Here ``tr'' is taken over the Dirac, flavor and color matrices. $V_1[S]$ 
corresponds to the 1-loop potential with the quark 1-loop diagram and
$V_2[S]$ is the 2-loop potential with the one gluon exchange. 
$\Sigma(p^2)$ has a general form;
\be
\Sigma(p) = \int \frac{d^4k}{(2\pi)^4} g \frac{\lambda^a}{2} \gmu
              S(k^2) g \Gamma^a_{\nu}(p,k) D_{\mu\nu}(p-k) ,
\label{eq:T0self}
\ee
where $\Gamma^a_{\nu}(p,k)$ and $D_{\mu\nu}(p-k)$ 
are the dressed quark-gluon vertex function and
the dressed gluon propagator respectively.

 Eq. (\ref{eq:T0sdeq}) can be decomposed into two coupled integral 
equations for $A(p^2)$ and $B(p^2)$
by taking the trace ``Tr'' over the gamma matrices
\cite{SDreview,SDdecomposition};
\beq
A(p^2) &=& 1 - \frac{i}{4p^2} \mbox{Tr}\left[  \slash \hspace{-5.8pt} p
         \Sigma(p) \right] ,
\label{eq:T0A} \\
B(p^2) &=& \frac{1}{4} \mbox{Tr}\left[ \Sigma(p) \right] .
\label{eq:T0B}
\eeq
Since $\Sigma(p)$ contains the dressed quark propagator, right-hand sides 
of eqs. (\ref{eq:T0A}) and (\ref{eq:T0B}) are implicit functions of 
$A(k^2)$ and $B(k^2)$ integrated over the momentum $k$.
Once one obtains the solutions $A(p^2)$ and $B(p^2)$, 
the effective potential corresponding to those solutions are given by
\beq
V[S] &=& -4 N_c \int \frac{d^4p}{(2\pi)^4} \left\{ \ln \left( p^2 A^2(p^2) +
       B^2(p^2) \right) + p^2 \sigma_A(p^2) - \ln p^2 - 1 \right\}
     \nonumber \\
     &=& - \frac{N_c}{\pi^3} \int^{\infty}_{-\infty} dp_4 
         \int^{\infty}_0 d|\p| \,\p^2 
         \left\{ \frac{}{} \right. \ln \left( (\p^2+p_4^2) A^2(\p^2,p_4^2) +
         B^2(\p^2,p_4^2) \right) 
     \nonumber \\
     && \hspace{120pt} + (\p^2+p_4^2) \sigma_A(\p^2,p_4^2) 
       - \ln( \p^2+p_4^2 ) - 1 \left. \frac{}{} \right\} .
\label{eq:T0eff}
\eeq 
To make the connection to $T \neq 0$ case clear, the integration
over the 4-momentum are decomposed into $|\p|$ and $p_4$ instead of 
$p^2 = \p^2 + p_4^2$ in eq. (\ref{eq:T0eff}). In eq. (\ref{eq:T0eff}) 
the factor $2$ for the degenerate ($u,d$)-quark is included
explicitly.

 In order to solve eqs. (\ref{eq:T0A}) and (\ref{eq:T0B}) in terms of
$A(\p^2,p_4^2)$ and $B(\p^2,p_4^2)$, one needs to define the structure
of the dressed gluon propagator $D_{\mu\nu}$, the dressed quark-gluon
vertex funciton $\Gamma_{\mu}^a$ and the coupling constant $g^2$. 
In this paper we adopt the following model: 
For the dressed quark-gluon vertex function, 
we use the rainbow approximation 
$\Gamma^a_{\mu}(p,k) = \frac{\lambda^a}{2}\gmu$, where $\lambda^a$ is 
the color Gell-Mann matrices. For the coupling constant $g^2$, the QCD 
running coupling constant in the one-loop order with an infrared
regulator $p_c$ is used to describe the momentum dependence 
\cite{Higashijima_2};
\be
\frac{g^2(p^2)}{4\pi} = \frac{4\pi}{9} \,
                        \frac{1}{\ln[(p^2+p^2_c)/\La^2]} \, .
\label{eq:QCDcoupling}
\ee
$p_c$ approximately divides the momentum scale into the
infrared region and the ultraviolet region. Furthermore, we adopt
the Higashijima-Miransky approximation 
\cite{Higashijima,Miransky}
for the QCD running coupling constant
\be
g^2\left( (p-k)^2 \right) = \theta(p^2-k^2)g^2(p^2) + \theta(k^2-p^2)g^2(k^2) 
         = g^2\left(\max[p^2,k^2]\right) \, .
\ee
In this approximation the quark mass-function $B(p^2)$ in the ultraviolet
region has an asymptotic form consistent with that obtained by the
operator product expansion and the renormalization group. 
For the dressed gluon propagator, we use the free 
one $D^{(0)}_{\mu\nu}(p-k)$ in the Landau-gauge,
\be
D^{(0)}_{\mu\nu}(p) = \left( \delta_{\mu\nu} - \frac{p_{\mu}p_{\nu}}{p^2}
                      \right) \frac{1}{p^2} \,.
\ee
The Landau-gauge is a fixed point of the renormalization group and 
the gauge parameter is zero to all orders in perturbation theory
\cite{Landau-gauge}.

 The integration over the angle between 
$p$ and $k$ can be performed in (\ref{eq:T0self}) because the running 
coupling constant has no angle dependence.
Then the quark self-energy has no term with odd number of $\gmu$,
that is Tr$[ \slash \hspace{-5.8pt} p \Sigma(p) ] = 0$.
As a consequence, eq. (\ref{eq:T0A})
reduces to $A(p^2) = 1$ in this model as mentioned in Sec.~I.
The SD equation for the quark becomes a single integral equation 
(\ref{eq:T0B}) for $B(p^2)$ after integrating over the angle between 
$p$ and $k$, 
\be
B(p^2) = \frac{3C_F}{16\pi^2} \int dk^2 \frac{g^2\left(\max[p^2,k^2]\right)}
         {\max[p^2,k^2]} \cdot \frac{k^2 B(k^2)}{k^2 + B^2(k^2)} \, ,
\label{eq:T0SDB}
\ee
where $C_F = (N_c^2-1)/2N_c$ is the quadratic Casimir operator for the 
color SU($N_c$) group. Eq. (\ref{eq:T0SDB}) can be expressed in term of 
the variables $\p$ and $p_4$ instead of 
$p^2 = \p^2 + p_4^2$,
\be
B(\p^2,p_4^2) = \int^{\infty}_{-\infty} dk_4 \int^{\infty}_0 d|\k|
                 \frac{\k^2 B(\k^2,k_4^2)}{\k^2+k_4^2+B^2(\k^2,k_4^2)}
                 \cdot E_1(|\p|,p_4;|\k|,k_4) \, .
\label{eq:T0sdeqB}
\ee
The explicit form of $E_1(|\p|,p_4;|\k|,k_4)$ is given in Appendix.
Eq. (\ref{eq:T0sdeqB}) has two distinct solutions; 
one is a trivial solution
$B_W(\p^2,p_4^2) = 0$ representing a free quark, and another is
a non-trivial solution $B_{NG}(\p^2,p_4^2) \neq 0$. 
The subscprit $W$ represents the ``Wigner'' solution describing a
phase in which the chiral symmetry is not broken. 
The subscprit $NG$ represents the ``Nambu-Goldstone'' (NG) solution which
represents a phase in which the chiral symmetry is broken spontaneously.
To determine the ture gound state realized at $T = 0$,
the effective potential $V[S]$ with the Wigner solution and that
with the NG solution have to be compared.

\subsection{SD eq. for the quark at $T \neq 0$}

 In this subsection, we extend the formulas in the previous subsection
to $T \neq 0$.  For this purpose, we apply the imaginary time 
formalism
\cite{ITF}
and make a replacement 
\be
\int \frac{d^4p}{(2\pi)^4} f(\p,p_4) \rightarrow
T \sum_{n=-\infty}^{\infty} \int \frac{d^3\p}{(2\pi)^3}
f(\p,\wn) \, ,
\label{eq:imaginary}
\ee 
where $\wn = (2n+1)\pi T \, (n \in Z)$ is the Matsubara-frequency for 
the fermion. The SD equation for the quark (\ref{eq:T0sdeq}) 
at $T \neq 0$ is decomposed into three coupled integral equations 
for $A(\p^2,\wn^2),C(\p^2,\wn^2)$ and $B(\p^2,\wn^2)$:
\beq
B(\p^2,\wn^2) &=& C_F \, T \sum_{m=-\infty}^{\infty} 
                  \int \frac{d^3\k}{(2\pi)^3}
                  g^2 \left(\max[\p^2+\wn^2,\k^2+\wm^2] \right) \nonumber \\
              & & \hspace{10pt} \times
                  \frac{B(\k^2,\wm^2)}{\k^2 A^2(\k^2,\wm^2) 
                  + \wm^2 C^2(\k^2,\wm^2)
                  + B^2(\k^2,\wm^2)} 
                  D^{(0)}_{\mu\mu}(\p-\k,\wn-\wm) \, ,
                  \label{eq:sdeqB}\\
A(\p^2,\wn^2) &=& 1 - \frac{C_F}{\p^2}\,\, T \sum_{m=-\infty}^{\infty} 
                  \int \frac{d^3\k}{(2\pi)^3} \cdot
                  \frac{g^2 \left(\max[\p^2+\wn^2,\k^2+\wm^2] \right)}
                  {\k^2 A^2(\k^2,\wm^2) + \wm^2 C^2(\k^2,\wm^2)
                  + B^2(\k^2,\wm^2)}  \nonumber \\
              & & \hspace{10pt} \times \left. \frac{}{} \right\{ 
                  A(\k^2,\wm^2) \left[ 2p_jk_i D^{(0)}_{ji}(\p-\k,\wn-\wm)
                  - \p \cdot \k D^{(0)}_{\mu\mu}(\p-\k,\wn-\wm) \right]
                  \nonumber \\
              & & \hspace{153pt} + \left.  2 C(\k^2,\wm^2) \wm p_j 
                  D^{(0)}_{j4}(\p-\k,\wn-\wm) \frac{}{} \right\} \, ,
                  \label{eq:sdeqA}\\
C(\p^2,\wn^2) &=& 1 - \frac{C_F}{\wn}\,\, T \sum_{m=-\infty}^{\infty} 
                  \int \frac{d^3\k}{(2\pi)^3} \cdot
                  \frac{g^2 \left(\max[\p^2+\wn^2,\k^2+\wm^2] \right)}
                  {\k^2 A^2(\k^2,\wm^2) + \wm^2 C^2(\k^2,\wm^2)
                  + B^2(\k^2,\wm^2)}  \nonumber \\
              & & \hspace{10pt} \times \left. \frac{}{} \right\{
                  2 A(\k^2,\wm^2) k_j 
                  D^{(0)}_{4i}(\p-\k,\wn-\wm) \nonumber \\
              & & \hspace{45pt} + \wm C(\k^2,\wm^2) 
                  \left[ 2 D^{(0)}_{44}(\p-\k,\wn-\wm)
                   - D^{(0)}_{\mu\mu}(\p-\k,\wn-\wm) \right] \left. 
                  \frac{}{} \right\} \, .
                  \label{eq:sdeqC}
\eeq
By making the integration over the angle between $\p$ and $\k$,
eqs. (\ref{eq:sdeqB}), (\ref{eq:sdeqA}) and (\ref{eq:sdeqC})
reduce to the following equations:
\beq
B(\p^2,\wn^2) &=& 2 \pi T \sum_{m=-\infty}^{\infty} 
                  \int d|\k| \frac{\k^2 B(\k^2,\wm^2)}{\k^2 A^2(\k^2,\wm^2) 
                  + \wm^2 C^2(\k^2,\wm^2)
                  + B^2(\k^2,\wm^2)} \cdot
                  E_1(|\p|,\wn;|\k|,\wm) \, , 
                  \label{eq:B}\\
A(\p^2,\wn^2) &=& 1 + \frac{2 \pi T}{\p^2} \sum_{m=-\infty}^{\infty} 
                  \int d|\k| \frac{\k^2}{\k^2 A^2(\k^2,\wm^2) 
                  + \wm^2 C^2(\k^2,\wm^2)
                  + B^2(\k^2,\wm^2)} \nonumber \\
              & & \hspace{70pt} \times \left\{
                  A(\k^2,\wm^2)E_2(|\p|,\wn;|\k|,\wm) + 
                  \wm C(\k^2,\wm^2)E_3(|\p|,\wn;|\k|,\wm) \right\} \, ,
                  \label{eq:A}\\
C(\p^2,\wn^2) &=& 1 + \frac{2 \pi T}{\wn} \sum_{m=-\infty}^{\infty} 
                  \int d|\k| \frac{\k^2}{\k^2 A^2(\k^2,\wm^2) 
                  + \wm^2 C^2(\k^2,\wm^2)
                  + B^2(\k^2,\wm^2)}  \nonumber \\
              & & \hspace{70pt} \times \left\{
                  A(\k^2,\wm^2)E_4(|\p|,\wn;|\k|,\wm) + 
                  \wm C(\k^2,\wm^2)E_5(|\p|,\wn;|\k|,\wm) \right\} \, .
                  \label{eq:C}
\eeq
Explicit forms of $E_i(|\p|,\wn;|\k|,\wm)\, (i=1,2,3,4,5)$
are given in Appendix. Eq. (\ref{eq:B}) may have two distinct
solutions: $B = B_W(\p^2,\wn^2) =0$ and $B = B_{NG}(\p^2,\wn^2) \neq 0$.
We call the solutions of eqs. (\ref{eq:A}) and (\ref{eq:C})
corresponding to $B_W$ as ($A_W,C_W$) and to $B_{NG}$ as 
($A_{NG},C_{NG}$).
The dressed quark propagators in the Wigner phase
and the NG phase are defined by
\beq
S_W(\p,\wn) &\equiv& \frac{1}{i \bg \cdot \p A_W(\p^2,\wn^2) 
                + i \gfour \wn C_W(\p^2,\wn^2)} \nonumber \\
            &\equiv& -i\bg \cdot \p \sigma_{A_W}(\p^2,\wn^2) - i \gfour \wn 
                     \sigma_{C_W}(\p^2,\wn^2) \, , \label{eq:T_Wigner}\\
S_{NG}(\p,\wn) &\equiv& \frac{1}{i \bg \cdot \p A_{NG}(\p^2,\wn^2) 
          + i \gfour \wn C_{NG}(\p^2,\wn^2) + B_{NG}(\p^2,\wn^2)} \nonumber \\
          &\equiv& -i\bg \cdot \p \sigma_{A_{NG}}(\p^2,\wn^2) - i \gfour \wn 
          \sigma_{C_{NG}}(\p^2,\wn^2) + \sigma_{B_{NG}}(\p^2,\wn^2).
\label{eq:T_NG}
\eeq
Unlike the case at $T = 0$, the quark self-energy $\Sigma(\p^2,\wn^2)$
can have terms with odd powers of $\gamma_{\mu}$ at $T \neq 0$.
Therefore, $A$ and $C$ cannot be equal to unity even when we take 
the rainbow approximation with the free gluon propagator in the 
Landau-gauge. Furthermore, $A$ and $C$ have non-trivial dependence
on $\p^2$ and $\wn^2$.
\footnote{In a simple algebraically soluble model, it was noted before
          that this non-trivial dependence has significant effect on
          the bulk thermodynamic quantities \cite{SDreview}.}

 Once one obtains $A, C$ and $B$, the effective potential at $T \neq 0$
as a direct generalization of eq. (\ref{eq:imaginary})
can be calculated through the formula,
\beq
V[S] &=& - \frac{2N_c}{\pi^2} \,\, 
       T \sum_{n=-\infty}^{\infty} \int d|\p| \, \p^2 
       \left. \frac{}{} \right\{ 
       \ln \left( \p^2 A^2(\p^2,\wn^2) + \wn^2 C^2(\p^2,\wn^2)
       + B^2(\p^2,\wn^2) \right) \nonumber \\
     & & \hspace{110pt} +  \p^2 \sigma_A(\p^2,\wn^2)
       + \wn^2 \sigma_C(\p^2,\wn^2) - \ln \left( \p^2 + \wn^2 \right) 
       - 1 \left. \frac{}{} \right\} .
\eeq
To study the chiral phase transition, the difference of the free energy
between the NG phase ($B_{NG} \neq 0$) and the Wigner phase ($B_W=0$)
should be considered:
\beq
\bar{V}(T) &\equiv& V[S_{NG}] - V[S_W]  \nonumber \\
     &=& - \frac{2N_c}{\pi^2} \,\, 
       T \sum_{n=-\infty}^{\infty} \int d|\p| \, \p^2 
       \left. \frac{}{} \right\{ 
       \ln \left[ \frac{\p^2 A_{NG}^2(\p^2,\wn^2) 
       + \wn^2 C_{NG}^2(\p^2,\wn^2)
       + B_{NG}^2(\p^2,\wn^2)}{\p^2 A_W^2(\p^2,\wn^2) 
       + \wn^2 C_W^2(\p^2,\wn^2)} \right] \nonumber \\
     & & \hspace{30pt} +  \p^2 \left[\sigma_{A_{NG}}(\p^2,\wn^2) 
       - \sigma_{A_W}(\p^2,\wn^2)\right]
       + \wn^2 \left[ \sigma_{C_{NG}}(\p^2,\wn^2) -
       \sigma_{C_W}(\p^2,\wn^2)\right]
       \left. \frac{}{} \right\} .
\label{eq:stability}
\eeq
If $\bar{V}(T) > 0 (<0)$, the chiral symmetry is restored (spontaneously
broken).

%%%%%%%%%%%%%%%%%%%%%%%%%%%%%%%%%%%%%%%%%%%%%%%%%%%%%%%%%%%%%%%%%%%%%%%%%%
%                         SECTION.3
%%%%%%%%%%%%%%%%%%%%%%%%%%%%%%%%%%%%%%%%%%%%%%%%%%%%%%%%%%%%%%%%%%%%%%%%%%

\section{CHIRAL PHASE TRANSITION AT FINITE TEMPERATURE}

\subsection{Determination of the parameters at $T = 0$}

 In this subsection, we determine the parameters of the model
in Sec.~II. A at $T = 0$. First of all, the momentum integrals
in eqs. (\ref{eq:T0eff}) and (\ref{eq:T0sdeqB}) are regulated
by the ultraviolet cutoff $\Lambda$. As far as $\Lambda$ is large
enough, physical quantities such as $f_{\pi}$ and $V[S]$ do not depend
on $\Lambda$, while the $\Lambda$-dependence of the chiral condensate
$\CC_{\Lambda}$ is governed by the renormalization group equation.
In our numerical calculation, we adopt $\Lambda = 5.42$ GeV above which
$f_{\pi}$ and $V[S]$ are insensitive to $\Lambda$.

 The chiral condensate $\CC_{\Lambda}$ is known to be insensitive to 
the infrared regularization parameter $p_c$
\cite{KMT}.
Therefore we take $p_c^2/\La^2=e^{0.1}$ and determine $\La$ to 
reproduce the pion decay constant $f_{\pi}=93$ MeV
\cite{KMT}.
As mentioned in Sec.~II.~A, eq. (\ref{eq:T0sdeqB}) has two distinct 
solutions. One is a trivial solution $B_W(\p^2,p_4^2) = 0$ corresponding
to the free quark. For this solution, $f_{\pi} = 0$ and 
$\CC_{\Lambda} = 0$. Another solution is $B_{NG}(\p^2,p_4^2)\neq 0$ 
which represents the spontaneous chiral symmetry
breaking. This solution is obtained by solving eq. (\ref{eq:T0sdeqB}) 
numerically by iteration, starting from an arbitrary given trial function
$B_{\mbox{\scriptsize trial}}(\p^2,p_4^2)\neq 0$. In this case, 
$f_{\pi}$ can be calculated from the formula
\cite{PDC},
\be
f_{\pi} N_{\pi} = \frac{N_c}{\pi^3} 
            \int^{\Lambda}_{-\Lambda} dp_4 \int^{\Lambda}_0 d|\p| \,\p^2
            B(\p^2,p_4^2) \left\{\sigma_A\sigma_B
            + \frac{2}{3} \p^2 \left(\sigma'_A
            \sigma_B + \sigma_A \sigma'_B \right) \right\} \, ,
\label{eq:f_pi}
\ee
with $\sigma'_B \equiv \del\sigma_B(\p^2,p_4^2)/\del\p^2$ and
$\sigma'_A \equiv \del\sigma_A(\p^2,p_4^2)/\del\p^2$.
$N_{\pi}$ is a canonical normalization constant for the 
Bethe-Salpeter amplitude in the ladder approximation,
\beq
N^2_{\pi} &=& \frac{N_c}{2\pi^3} \left. \int^{\Lambda}_{-\Lambda} 
       dp_4 \int^{\Lambda}_0 d|\p| \,\p^2
       B^2(\p^2,p_4^2) \right\{ \sigma^2_A -2 \left[ \p^2 \sigma_A \sigma'_A + 
       p_4^2 \sigma_C \sigma'_C + \sigma_B \sigma'_B \right]
       \nonumber \\
       && \hspace{20pt} \left. - \frac{4}{3}\p^2 \left[ \frac{}{} \right.
            \p^2\left(\sigma_A \sigma''_A
          - (\sigma'_A)^2 \right) + p_4^2\left(\sigma_C \sigma''_C
          - (\sigma'_C)^2 \right) + \sigma_B \sigma''_B
          - (\sigma'_B)^2
         \left. \frac{}{} \right] \right\}  \, ,
\label{eq:normal}
\eeq
with $\sigma'_C \equiv \del\sigma_C(\p^2,p_4^2)/\del\p^2 =
\sigma'_A$ at $T = 0$. If $A = C = 1$, one obtains $N_{\pi} = f_{\pi}$
\cite{PDC}, and
$\La = 734$ MeV is determined by the input
$f_{\pi} = 93$ MeV in eqs. (\ref{eq:f_pi}) or (\ref{eq:normal})
with $A = C = 1$ and $B_{NG}$.
Our parameters are consitent with those obtained in 
\cite{KMT}. 

 The effective potential for the solution $B_{NG}$ is 
obtained from eq. (\ref{eq:T0eff}). The difference of the vacuum
energy between the NG phase and the Wigner phase reads
$\bar{V}(0) = - (172.7\mbox{ MeV})^4 < 0$.
This means that the state corresponding to 
the NG solution is realized at $T = 0$. 
At the scale $\Lambda$, the chiral condensate $\CC_{\Lambda}$ 
is calculated
as
\beq
\CC_{\Lambda} &=& - \int^{\Lambda} \frac{d^4p}{(2\pi)^4} \mbox{tr}[S(p)]
              \nonumber \\
              &=& - \frac{3}{\pi^3} 
                \int^{\Lambda}_{-\Lambda} dp_4 \int^{\Lambda}_0 d|\p| \,\p^2
                \sigma_B(\p^2,p_4^2) \, .
\label{eq:T0CC}
\eeq
With the parameter set determined above, we obtain 
$\CC_{\Lambda} = (-293\mbox{ MeV})^3$. If we change the scale from 
$\Lambda$ to 1 GeV by the perturbative renormalization group equation,
$\CC_{1 \mbox{\scriptsize GeV}} = (-217\mbox{ MeV})^3$ 
is obtained, which is consistent
with the known phenomenological value.

\subsection{The exact numerical solutions of the SD eq. in the rainbow
            approximation at $T \neq 0$}

 At $T \neq 0$, we need to replace the $dp_4$ 
integral by the Matsubara sum,
\be
\int^{\Lambda}_{-\Lambda} \frac{dp_4}{2 \pi} f(\p,p_4)\rightarrow 
     T \sum_{n=-n_{\mbox{\tiny max}}}^{n_{\mbox{\tiny max}}} f(\p,\wn)\, ,
\label{eq:replace} 
\ee
with $n_{\mbox{\tiny max}}$ being the largest integer $n$ satisfying 
$\wn = (2n + 1)\pi T \le \Lambda$.
In order to study how the quark wave-function renormalization affects
the nature of the chiral phase transition, eqs. (\ref{eq:B}), 
(\ref{eq:A}) and (\ref{eq:C}) are solved in two different cases
with the parameter set determined at $T = 0$.
\begin{itemize}
\item[(I)$\,\,$]
: $A$ and $C$ are assumed to be $1$ and only eq. (\ref{eq:B}) for $B$ is 
solved numerically. The trivial solution $B = 0$ of eq. 
(\ref{eq:B}) corresponds to the free quark and the numerical solution
$B_{NG} \neq 0$ corresponds to the phase of the spontaneous 
chiral symmetry breaking. We define the difference of the free energy
between the NG phase and the Wigner phase in this case as 
$\bar{V}_{(I)}(T)$ according to eq. (\ref{eq:stability}) with
$A_W = C_W = A_{NG} = C_{NG} = 1$.

\item[(II)$\,$]
: Three coupled eqs. (\ref{eq:B}), (\ref{eq:A}) and (\ref{eq:C}) are
solved numerically with initial trial functions for $A, B$ and $C$.
We call the solutions corresponding to the Wigner phase as 
$B_W (= 0), A_W (\neq 1)$ and $C_W (\neq 1)$ while those corresponding
to the NG phase as $B_{NG} (\neq 0), A_{NG} (\neq 1)$ and 
$C_{NG} (\neq 1)$. The difference of the free energy between the NG 
phase and the Wigner phase is defined as $\bar{V}_{(II)}(T)$
according to eq. (\ref{eq:stability}).
\end{itemize}

 Fig.~1 shows the momentum dependence of the quark mass-function 
$B_{NG}(\p^2,\omega_0^2)$ in the case (I) with a zeroth 
Matsubara-frequency $\omega_0 = \pi T$  at $T=10,150,200$ and $216$
MeV. $B_{NG}(\p^2,\wn^2)$ in the 
low momentum region decreases with the increase of $|n|$ at given
$T$. For $T \geq 217$ MeV, eq. (\ref{eq:B}) have only the 
trivial solution $B_W=0$.

 Fig.~2 (a) and (b) show the momentum dependence of the functions 
$A_W$ and $C_W$ in the case (II) with $\omega_0 = \pi T$. It is found
that $A_W$ and $C_W$ in the low momentum region do not reach unity
even at $T \sim$ a few hundred MeV. This implies that the quarks
and gluons are strongly interacting even in the high $T$ chirally
symmetric phase. For $\wn (n \neq 0)$, the difference of $A_W$ and 
$C_W$ from unity decreases with the increase of $|n|$ at given $T$.

 The momentum dependence of the quark mass-function $B_{NG}$ in the case
(II) with $\omega_0 = \pi T$ is shown in Fig.~3. The non-trivial
solution $B_{NG}\neq 0$ is obtained only below $T = 155$ MeV 
and eq. (\ref{eq:B}) have only a trivial solution $B_W=0$
for $T \geq 155$ MeV. $B_{NG}$ for $\wn (n \neq 0)$ in the low momentum
region becomes smaller as $|n|$ increases at given $T$.

 Fig.~4 (a) and (b) show the momentum dependence of the 
functions $A_{NG}$ and $C_{NG}$ with $\omega_0 = \pi T$. 
It is found that the difference of $A_{NG}$ and 
$C_{NG}$ from unity in the low momentum region
is maximal around $T=150$ MeV, and
then $A_{NG}$ and $C_{NG}$ coincide with $A_W$ and $C_W$ 
for $T \geq 155$ MeV. For $\wn (n \neq 0)$,
the difference of $A_{NG}$ and $C_{NG}$ from unity decreases
with the increase of $|n|$ at given $T$.

 As mentioned before, in order to determine the critical temperature 
$T_c$ of the chiral symmetry restoration, one has to calculate the 
difference of the effective potential between the NG phase and the
Wigner phase, $\bar{V}(T)$ in eq. (\ref{eq:stability}). 
$T_c$ is defined by the temperature at which $\bar{V}(T_c) = 0$. 
Fig.~5 shows the $T$-dependence of $\bar{V}_{(I)}(T)$ and 
$\bar{V}_{(II)}(T)$. It is found that $T_c = 217$ MeV in the case 
(I) and $T_c = 155$ MeV in the case (II), and the chiral phase
transition is of the second order in both cases. Due to the crude
ansatz $A = C = 1$ at $T \neq 0$ in the case (I), $\bar{V}_{(I)}(T)$
has a strange behavior in which it decreases first and then increases
as $T$ increases. Such behavior disappears in $\bar{V}_{(II)}(T)$
where $A$ and $C$ are solved together with $B$. $\bar{V}_{(II)}(T)$
increases monotonically as $T$ increases and vanishes at $T_c$.

 The pion decay constant $f_{\pi}$ at $T \neq 0$ is obtained
from eqs. (\ref{eq:f_pi}) and (\ref{eq:normal}) using eq. 
(\ref{eq:replace}). Fig.~6 shows the $T$-dependence of $f_{\pi}$
for the NG solution in the case (I) and (II).  In the case (I),
$f_{\pi}$ has a strange behavior in which it increases first and then
decreases as $T$ increases because of the crude approximation
$A = C = 1$. This behavior is correlated with that of 
$\bar{V}_{(I)}(T)$ in Fig.~5. On the other hand, $f_{\pi}$ 
in the case (II) decreases
slowly first and then changes suddenly near $T_c$ as $T$ increases.
The chiral condensate $\CC_{\Lambda}$ at the scale $\Lambda$ are 
calculated by eq.(\ref{eq:T0CC}) using eq.(\ref{eq:replace})
at $T \neq 0$. In Fig.~7, the $T$-dependence of $\CC_{\Lambda}$ for 
the NG solution in the case (I) and (II) is shown. 
For the same reason as that in $f_{\pi}$, 
the strange behavior of $\CC_{\Lambda}$ 
in the case (I) is found, while $\CC_{\Lambda}$ in the case (II) 
has an almost constant value first and 
then decreases suddenly near $T_c$ as $T$ increases.

 As mentioned above, the phase transtions in (I) and in (II) 
are both of the second order, so the critical exponents can be defined. 
Fig.~8 shows the scaling behaviour of $\CC_{\Lambda}$ in (I) and in (II) 
near the critical temperature in each case. 
The critical exponent $\beta$ is defined by
\be
 \CC_{\Lambda} \sim \left(1 - \frac{T}{T_c}\right)^{\beta} 
 \quad \mbox{for}
 \quad T \rightarrow T_c - 0\, . 
\ee
$\beta$ is determined by using the linear-log fit numerically
as follows
\be
\ln \CC_{\Lambda} = \beta \ln \left(1 - \frac{T}{T_c}\right) + 
                     \mbox{const.} \, ,
\ee
where const. is independent of $T$. We find that
$\beta$ $\sim 0.51$ in (I) and $\sim 0.53$ in (II) respectively. 
Both values are consistent with $\beta$ in the mean-field theory 
and agrees with ref.
\cite{KMT,SDcritical_expo}.
The mean-field critical exponents are a result
of the structure of the gap equation for the fermion.
The inclusion of the soft collective modes will move the mean field
exponents to the $O(4)$ values as expected from the renormalization
group argument
\cite{phase}.

%%%%%%%%%%%%%%%%%%%%%%%%%%%%%%%%%%%%%%%%%%%%%%%%%%%%%%%%%%%%%%%%%%%%%%%%%%
%                         SECTION.4
%%%%%%%%%%%%%%%%%%%%%%%%%%%%%%%%%%%%%%%%%%%%%%%%%%%%%%%%%%%%%%%%%%%%%%%%%%

\section{CONCLUSIONS AND DISCUSSIONS}

 In this paper, we have obtained the dressed quark propagator as the
numerical solutions by solving the SD eq. for the quark without any 
ansatz and studied the effect of the 
quark wave-function renormalization on the chiral phase transition 
at $T \neq 0$. The SD eq. for the quark has been solved at one-loop level.
We use the free gluon propagator in the Landau-gauge, the bare quark-gluon 
vertex, the one-loop QCD running coupling with the infrared regulator and 
Higashijima-Miransky approximation. 
The three-momentum and Matsubara-frequency dependent functions for the quark 
wave-funcition renormalization differ from unity especially in the low
momentum region. As a result the critical temperature decreases
substantially compared to that with $A = C = 1$.

 To examine the effect of the quark
wave-function renormalization, we have solved the SD eq. for the quark 
in two cases:
\begin{itemize}
\item[(I)$\,\,$]
: the SD eq. for $B$ is solved with an assumption $A = C = 1$.

\item[(II)$\,$]
: coupled SD eq. for $A,B$ and $C$ are solved numerically.
\end{itemize}
(I) is an approximate solution, and (II) is the exact
solutions of the SD eq. for the quark. The Wigner solution 
in the case (I) corresponds to the free quark.
From the difference of the free energy between 
the NG phase and the Wigner phase in the case (I),
the critical temperature of the chiral phase transition 
turns out to be $T_c =217$ MeV. 
Also the difference of the free energy between 
the NG phase and the Wigner phase in the case (II) 
shows $T_c = 155$ MeV. Exact numerical solution with 
$A \neq 1 \neq C$ in the case (II) 
not only reduces the critical temperature
but also removes the pathological behavior of $f_{\pi}$ and 
$\CC$ in the case (I).
We found that after the chiral symmetry
restoration, quarks are not free even at temperature of a few hundred
MeV in the case (II).
In both cases (I) and (II), 
the chiral phase transition is of the second order. 
The critical exponent $\beta$ is extracted from $\CC$ near $T_c$
and we obtained $\beta \sim 0.51$ in (I) and $\beta \sim 0.53$ in
(II). They are consistent with the 
result of 
\cite{SDreview,KMT,SDcritical_expo}.

 Finally, several comments are in order. 
The gluon propagator used in this paper is the free one and cannot
explain the quark and gluon confinement. To study the
effect of the quark confinement on the chiral phase transition,
it would be necessary to use the phenomenological gluon propagator 
such as the dual Ginzburg-Landau model which explains
the spontaneous chiral symmetry breaking and the quark confinement
\cite{KANAZAWA,RCNP}.
The screening of the gluon should be also included. The electric 
sector of the gluon propagator is influenced by the Debye screening
and the magnetic sector by the non-perturbative magnetic mass 
in the static case or the dynamical screening in the non static case.
The study of the chiral phase transition at finite density
and temperature is also an important future problem in our approach to
locate the triciritical point in the QCD phase diagram
\cite{Stephanov:1998dy}. Our preliminary study shows
that the location of the tricritical point in our model is
$(T_{tr},\mu^B_{tr}) = (210 \mbox{ MeV}, 129 \mbox{ MeV})$
for the case (I) and $(142 \mbox{ MeV}, 246 \mbox{ MeV})$ for the
case (II), where $\mu^B_{tr} (= 3 \mu_{tr})$ is the baryon chemical
potential.

%%%%%%%%%%%%%%%%%%%%%%%%%%%%%%%%%%%%%%%%%%%%%%%%%%%%%%%%%%%%%%%%%%%%%%%%%%
%                         ACKNOWLEDGEMENTS
%%%%%%%%%%%%%%%%%%%%%%%%%%%%%%%%%%%%%%%%%%%%%%%%%%%%%%%%%%%%%%%%%%%%%%%%%%

\section*{ACKNOWLEDGEMENTS}

 I am grateful to T. Hatsuda and S. Sasaki for their stimulating 
discussions, comments and encouragement.

%%%%%%%%%%%%%%%%%%%%%%%%%%%%%%%%%%%%%%%%%%%%%%%%%%%%%%%%%%%%%%%%%%%%%%%%%%
%                         APPENDIX
%%%%%%%%%%%%%%%%%%%%%%%%%%%%%%%%%%%%%%%%%%%%%%%%%%%%%%%%%%%%%%%%%%%%%%%%%%
\appendix
\section*{}
%{EXPLICIT FORMS OF FIVE KERNELS IN QUARK SD EQUATION}
 In this appendix, we show explicit forms of five kernels 
$E_i(|\p|,\wn;|\k|,\wm)\, (i=1,2,3,4,5)$
in eqs. (\ref{eq:T0B}), (\ref{eq:B}), (\ref{eq:A}) and (\ref{eq:C}):
\beq
E_1(|\p|,\wn;|\k|,\wm) &=&
    - \frac{4}{9\pi|\p||\k|} \cdot 
      \frac{1}{\ln \left[ \left(
      \max[\p^2+\wn^2,\k^2+\wm^2]+p^2_c \right)/\La^2 \right]} 
      \nonumber \\
      && \times
      \ln \frac{(|\p|-|\k|)^2+(\wn-\wm)^2}{(|\p|+|\k|)^2+(\wn-\wm)^2} \\
E_2(|\p|,\wn;|\k|,\wm) &=& 
    - \frac{8}{27\pi} \cdot 
      \frac{1}{\ln \left[ \left(
      \max[\p^2+\wn^2,\k^2+\wm^2]+p^2_c \right)/\La^2 \right]} 
      \nonumber \\
      && \times \left\{ 2 + \frac{\p^2+\k^2+3(\wn-\wm)^2}{4|\p||\k|} \cdot
      \ln \frac{(|\p|-|\k|)^2+(\wn-\wm)^2}{(|\p|+|\k|)^2+(\wn-\wm)^2} \right.
      \nonumber \\
      && \hspace{15pt} + \frac{ \left( \p^2-\k^2+(\wn-\wm)^2 \right)
               \left(-\p^2+\k^2+(\wn-\wm)^2 \right)}{4|\p||\k|}
      \nonumber \\
      && \hspace{25pt} \times 
      \left. \left( \frac{1}{(|\p|-|\k|)^2+(\wn-\wm)^2} -
      \frac{1}{(|\p|+|\k|)^2+(\wn-\wm)^2} \right) \right\} \\
E_3(|\p|,\wn;|\k|,\wm) &=& 
    - \frac{4(\wn-\wm)}{27\pi|\p||\k|} \cdot 
      \frac{1}{\ln \left[ \left(
      \max[\p^2+\wn^2,\k^2+\wm^2]+p^2_c \right)/\La^2 \right]} 
      \nonumber \\
      && \times \left\{ 
      \ln \frac{(|\p|-|\k|)^2+(\wn-\wm)^2}{(|\p|+|\k|)^2+(\wn-\wm)^2}
      \right. + \left(-\p^2+\k^2+(\wn-\wm)^2 \right)
      \nonumber \\
      && \hspace{15pt} \times 
      \left. \left( \frac{1}{(|\p|-|\k|)^2+(\wn-\wm)^2} -
      \frac{1}{(|\p|+|\k|)^2+(\wn-\wm)^2} \right) \right\} \\
E_4(|\p|,\wn;|\k|,\wm) &=& 
      \frac{4(\wn-\wm)}{27\pi|\p||\k|} \cdot 
      \frac{1}{\ln \left[ \left(
      \max[\p^2+\wn^2,\k^2+\wm^2]+p^2_c \right)/\La^2 \right]} 
      \nonumber \\
      && \times \left\{ 
      \ln \frac{(|\p|-|\k|)^2+(\wn-\wm)^2}{(|\p|+|\k|)^2+(\wn-\wm)^2}
      \right. + \left(\p^2-\k^2+(\wn-\wm)^2 \right)
      \nonumber \\
      && \hspace{15pt} \times 
      \left. \left( \frac{1}{(|\p|-|\k|)^2+(\wn-\wm)^2} -
      \frac{1}{(|\p|+|\k|)^2+(\wn-\wm)^2} \right) \right\} \\
E_5(|\p|,\wn;|\k|,\wm) &=& 
      \frac{4}{27\pi|\p||\k|} \cdot 
      \frac{1}{\ln \left[ \left(
      \max[\p^2+\wn^2,\k^2+\wm^2]+p^2_c \right)/\La^2 \right]} 
      \nonumber \\
      && \times \left\{ - 
      \ln \frac{(|\p|-|\k|)^2+(\wn-\wm)^2}{(|\p|+|\k|)^2+(\wn-\wm)^2}
      \right. + 2 (\wn-\wm)^2 
      \nonumber \\
      && \hspace{15pt} \times 
      \left. \left( \frac{1}{(|\p|-|\k|)^2+(\wn-\wm)^2} -
      \frac{1}{(|\p|+|\k|)^2+(\wn-\wm)^2} \right) \right\}
\eeq

%%%%%%%%%%%%%%%%%%%%%%%%%%%%%%%%%%%%%%%%%%%%%%%%%%%%%%%%%%%%%%%%%%%%%%%%%%
%                           REFERENCES
%%%%%%%%%%%%%%%%%%%%%%%%%%%%%%%%%%%%%%%%%%%%%%%%%%%%%%%%%%%%%%%%%%%%%%%%%%

\newpage

%%%%%%%%%%%%%%%%%%%%%%%%%%   FIG 1   %%%%%%%%%%%%%%%%%%%%%%%%%%%%%%%%
\begin{figure}[htbp]
    \centerline{
      \epsfxsize=0.49\textwidth
      \epsfbox{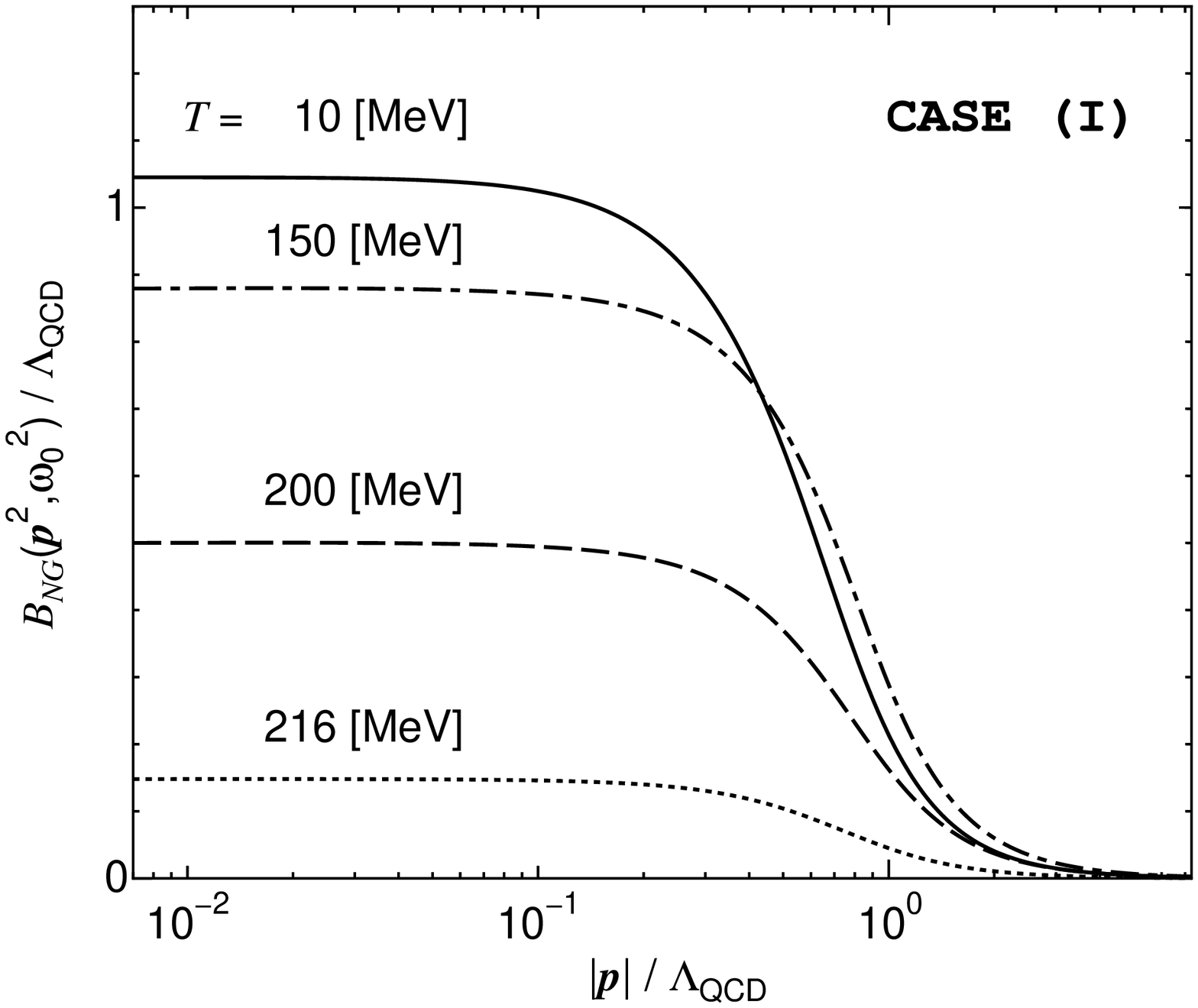}
    }
  \caption{The momentum dependence of the quark mass-function
           $B_{NG}(\p^2,\omega_0^2)$ in the case (I) at 
           $T=$10,150,200 and 216 MeV with the lowest
           Matsubara-frequency $\omega_0 = \pi T$.}
  \label{fig:1}
\end{figure}
%%%%%%%%%%%%%%%%%%%%%%%%%%%%%%%%%%%%%%%%%%%%%%%%%%%%%%%%%%%%%%%%%%%%%

%%%%%%%%%%%%%%%%%%%%%%%%%%   FIG 2   %%%%%%%%%%%%%%%%%%%%%%%%%%%%%%%%
\begin{figure}[htbp]
    \centerline{
      \epsfxsize=0.49\textwidth
      \epsfbox{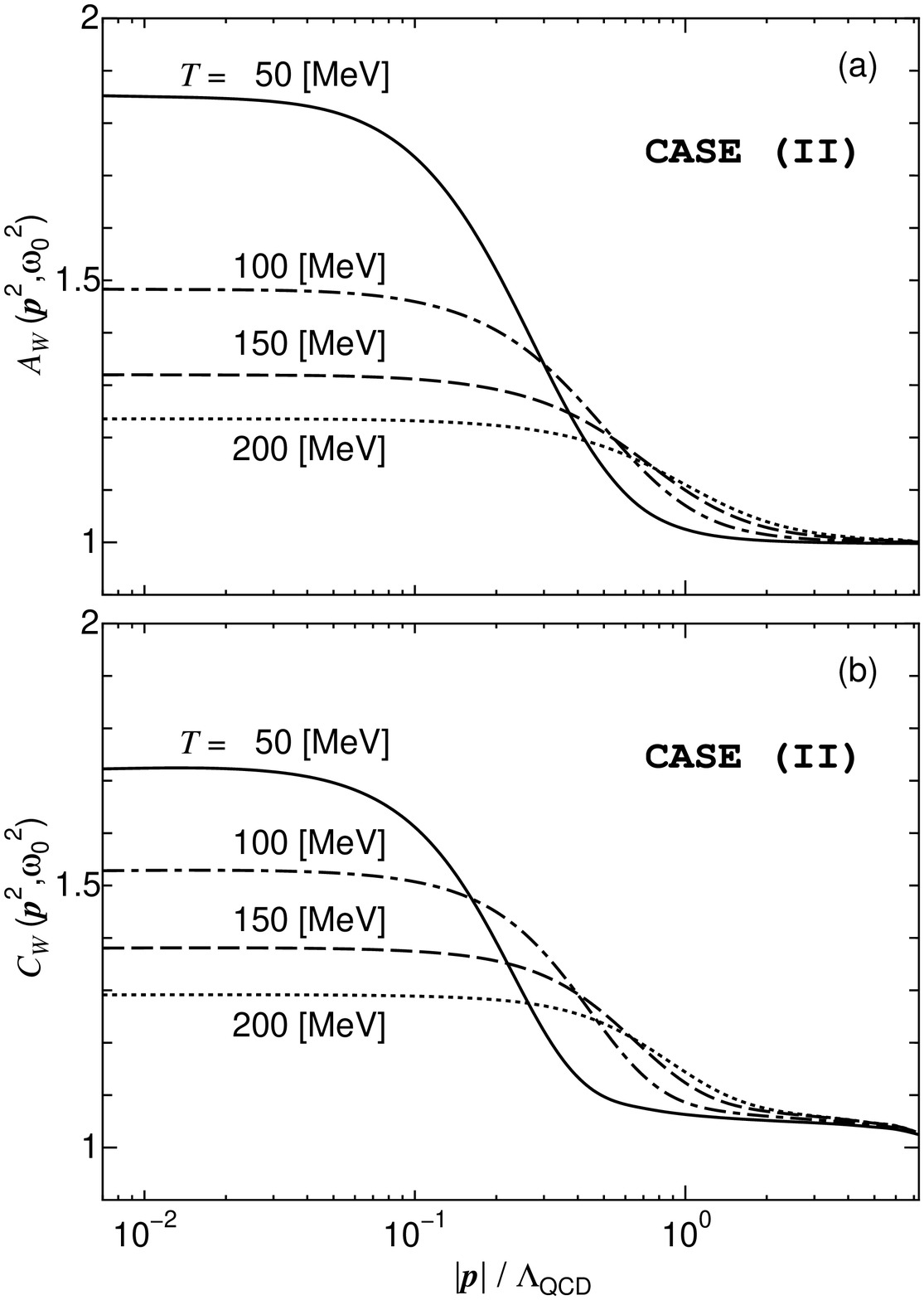}
    }
  \caption{The momentum dependence of the functions 
           $A_W(\p^2,\omega_0^2)$;(a) and $C_W(\p^2,\omega_0^2)$;(b)
           in the case 
           (II) at $T=$50,100,150 and 200 MeV
           with the lowest
           Matsubara-frequency $\omega_0 = \pi T$.}
  \label{fig:2}
\end{figure}
%%%%%%%%%%%%%%%%%%%%%%%%%%%%%%%%%%%%%%%%%%%%%%%%%%%%%%%%%%%%%%%%%%%%%

%%%%%%%%%%%%%%%%%%%%%%%%%%   FIG 3   %%%%%%%%%%%%%%%%%%%%%%%%%%%%%%%%
\begin{figure}[htbp]
    \centerline{
      \epsfxsize=0.49\textwidth
      \epsfbox{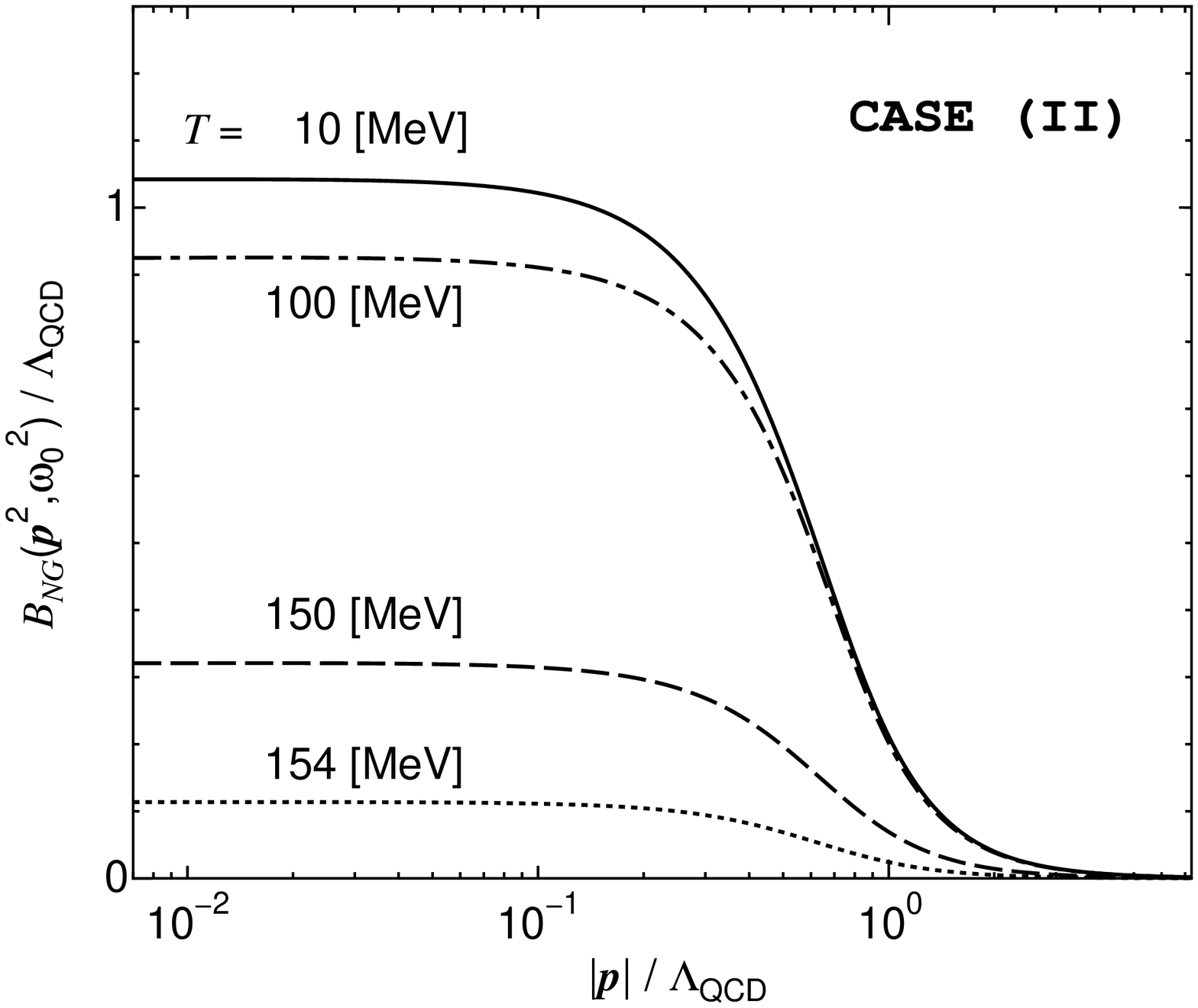}
    }
  \caption{The momentum dependence of the quark mass-function
           $B_{NG}(\p^2,\omega_0^2)$ in the case (II) at 
           $T=$10,100,150 and 154 MeV with the lowest
           Matsubara-frequency $\omega_0 = \pi T$.}
  \label{fig:3}
\end{figure}
%%%%%%%%%%%%%%%%%%%%%%%%%%%%%%%%%%%%%%%%%%%%%%%%%%%%%%%%%%%%%%%%%%%%%

%%%%%%%%%%%%%%%%%%%%%%%%%%   FIG 4   %%%%%%%%%%%%%%%%%%%%%%%%%%%%%%%%
\begin{figure}[htbp]
    \centerline{
      \epsfxsize=0.49\textwidth
      \epsfbox{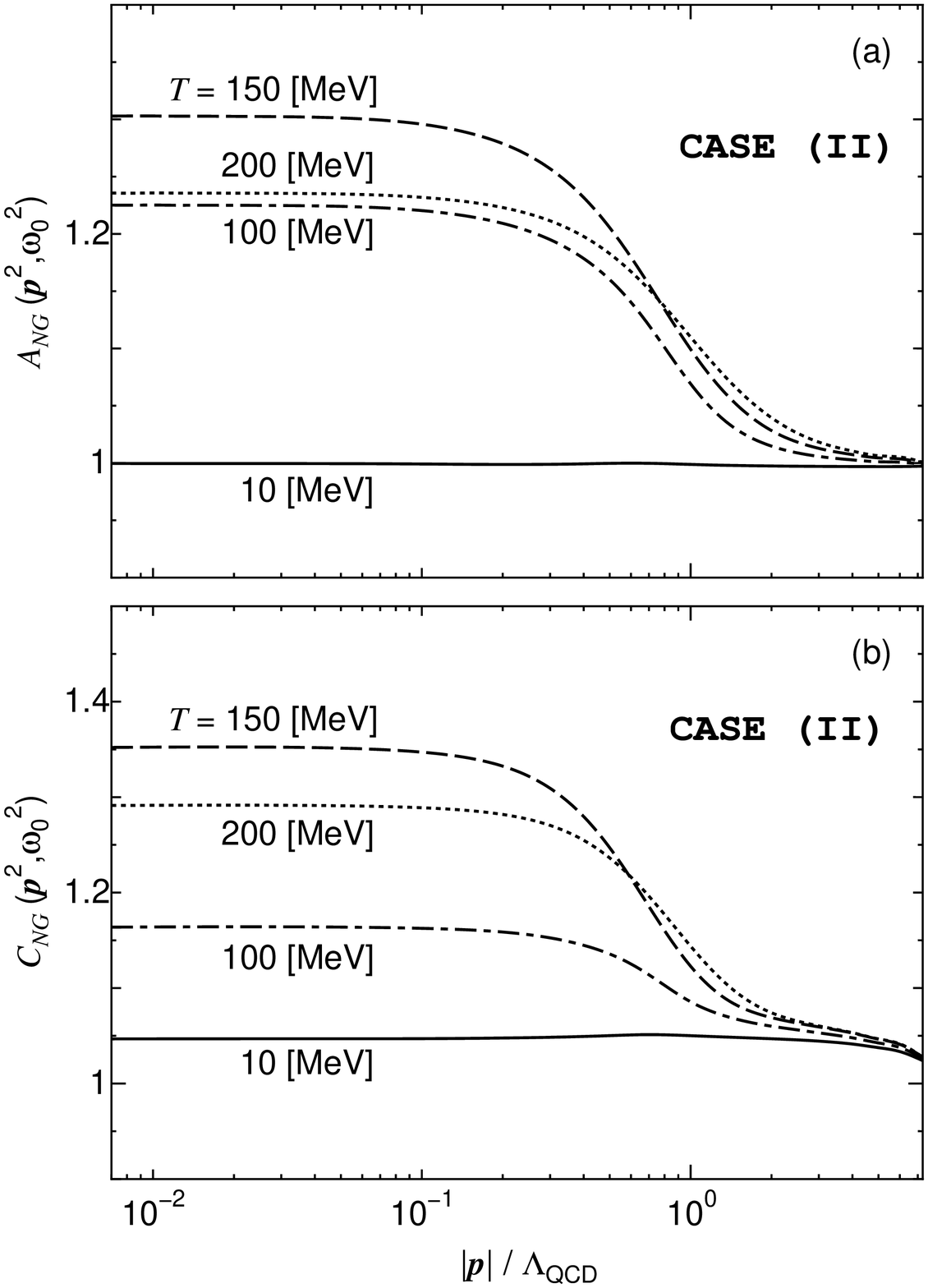}
    }
  \caption{The momentum dependence of the functions 
           $A_{NG}(\p^2,\omega_0^2)$;(a) and $C_{NG}(\p^2,\omega_0^2)$;(b)
           in the case 
           (II) at $T=$10,100,150 and 200 MeV
           with the lowest
           Matsubara-frequency $\omega_0 = \pi T$.}
  \label{fig:4}
\end{figure}
%%%%%%%%%%%%%%%%%%%%%%%%%%%%%%%%%%%%%%%%%%%%%%%%%%%%%%%%%%%%%%%%%%%%%

%%%%%%%%%%%%%%%%%%%%%%%%%%   FIG 5   %%%%%%%%%%%%%%%%%%%%%%%%%%%%%%%%
\begin{figure}[htbp]
    \centerline{
      \epsfxsize=0.49\textwidth
      \epsfbox{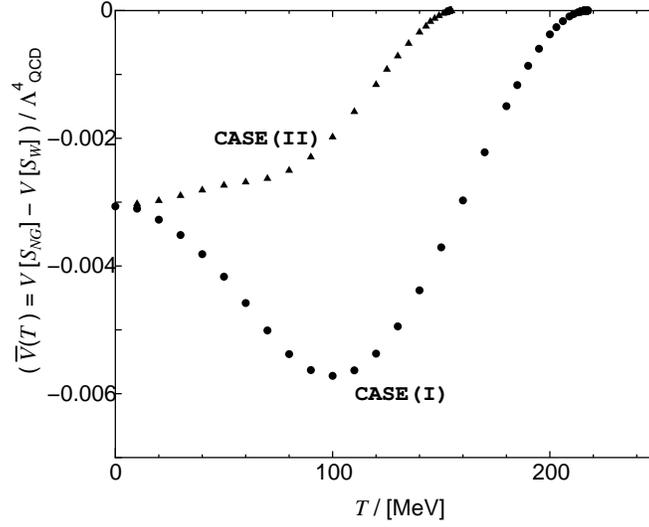}
    }
  \caption{The temperature dependence of the difference of the effctive 
           potential between the NG solution and the Wigner
           solution. Filled circles show $\bar{V}_{(I)}(T)$
           and filled triangles $\bar{V}_{(II)}(T)$.}
  \label{fig:5}
\end{figure}
%%%%%%%%%%%%%%%%%%%%%%%%%%%%%%%%%%%%%%%%%%%%%%%%%%%%%%%%%%%%%%%%%%%%%

%%%%%%%%%%%%%%%%%%%%%%%%%%   FIG 6   %%%%%%%%%%%%%%%%%%%%%%%%%%%%%%%%
\begin{figure}[htbp]
    \centerline{
      \epsfxsize=0.49\textwidth
      \epsfbox{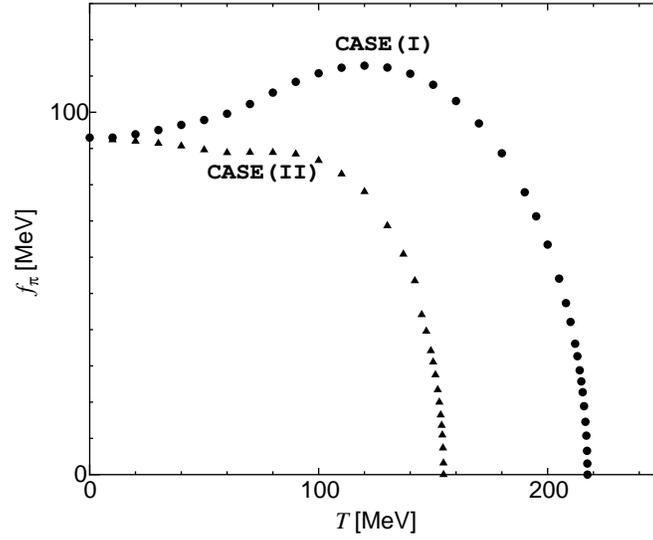}
    }
  \caption{The temperature dependence of the pion decay constant
           $f_{\pi}$ for the NG solutions in the case (I);(filled circles)
           and (II);(filled triangles).}
  \label{fig:6}
\end{figure}
%%%%%%%%%%%%%%%%%%%%%%%%%%%%%%%%%%%%%%%%%%%%%%%%%%%%%%%%%%%%%%%%%%%%%

%%%%%%%%%%%%%%%%%%%%%%%%%%   FIG 7   %%%%%%%%%%%%%%%%%%%%%%%%%%%%%%%%
\begin{figure}[htbp]
    \centerline{
      \epsfxsize=0.49\textwidth
      \epsfbox{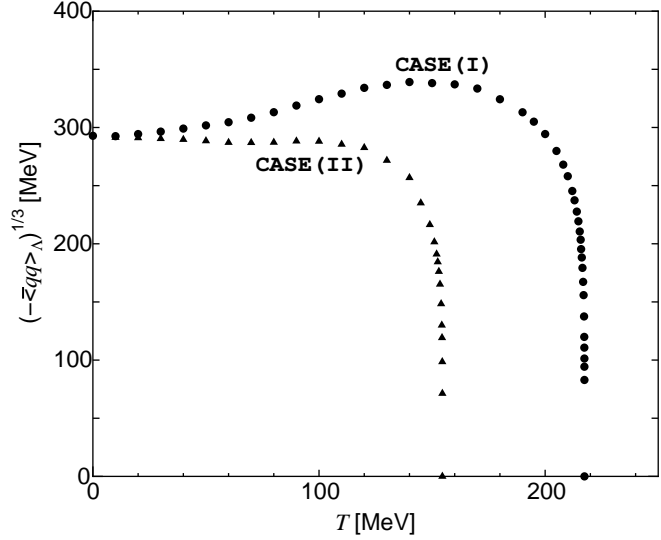}
    }
  \caption{The temperature dependence of the chiral condensate
           for the NG solutions in the case (I);(filled circles) 
           and (II);(filled triangles).}
  \label{fig:7}
\end{figure}
%%%%%%%%%%%%%%%%%%%%%%%%%%%%%%%%%%%%%%%%%%%%%%%%%%%%%%%%%%%%%%%%%%%%%

%%%%%%%%%%%%%%%%%%%%%%%%%%   FIG 8   %%%%%%%%%%%%%%%%%%%%%%%%%%%%%%%%
\begin{figure}[htbp]
    \centerline{
      \epsfxsize=0.49\textwidth
      \epsfbox{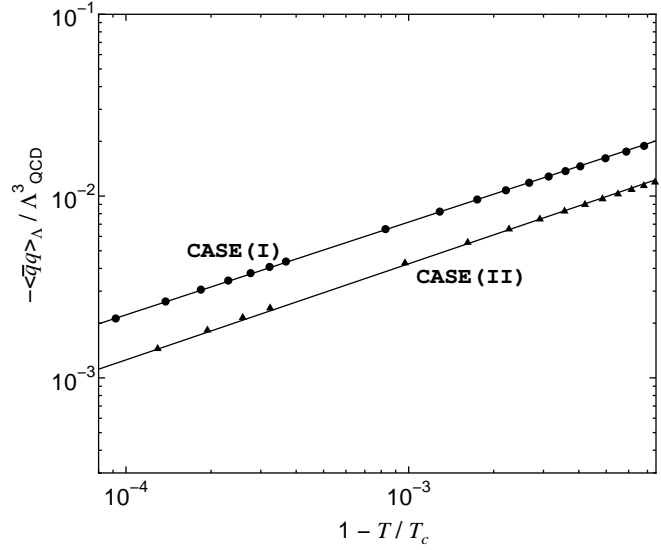}
    }
  \caption{The temperature dependence of the chiral condensate
           near the critical temperature $T_c$ for 
           the NG solutions in the case (I);(filled circles)  
           and (II);(filled triangles). 
           $T_c=$217 MeV in (I) and $T_c$=155 MeV in (II).}
  \label{fig:8}
\end{figure}
%%%%%%%%%%%%%%%%%%%%%%%%%%%%%%%%%%%%%%%%%%%%%%%%%%%%%%%%%%%%%%%%%%%%%

\end{document}